\documentclass[%
 aip,
 jmp,%
 amsmath,amssymb,
 reprint,%
]{revtex4-1}

\usepackage{epsfig}
\usepackage{graphics}
\usepackage{graphicx}
\usepackage{dcolumn}
\usepackage{bm}
\begin{document}

\preprint{AIP/123-QED}

\title{Hydrodynamic study of edge spin-vortex excitations of fractional quantum Hall fluid}

\author{M. Rabiu}
\email{mrabiu@uds.edu.gh}
\affiliation{University for Development Studies, Faculty of Applied Sciences, Department of Applied Physics, Navrongo Campus, Ghana.
}

\author{S. Y. Mensah}%
\affiliation{University of Cape Coast, Department of Physics, Laser and Fiber Optics Center, Cape Coast, Ghana.
}%

\author{I. Y. Seini}
 \affiliation{University for Development Studies, School of Engineering, Department of Mechanical Engineering, Nyankpala Campus, Ghana.
}

\author{S. S. Abukari}%
\affiliation{University of Cape Coast, Department of Physics, College of Physical Sciences, Cape Coast, Ghana.
}%

\date{\today}

\begin{abstract}
We undertake a theoretical study of edge spin-vortex excitations in fractional quantum Hall fluid. This is done in view of quantised Euler hydrodynamics theory. The dispersions of true excitations for fractions within $0\leq \nu \leq 1$ are simulated which exhibit universal similarities and differences in behaviour. The differences arise from different edge smoothness and spin (pseudo-spin) polarisations, in addition to spin-charge competition. In particular, tuning the spin-charge factor causes coherent spin flipping associated with partial and total polarisations of edge spin-vortices. This observation is tipped as an ideal mechanism for realisation of functional spintronic devices.
\end{abstract}

\maketitle

\section{Introduction}\label{Sec:Section1}
Two dimensional electron gas system submitted to strong magnetic field usually form strongly correlated quantum Hall fluids. The fluids are often incompressible \cite{Laughlin1983} and characterised by dissipationless superfluid flow \cite{PAlexandr2012} and formation of quantised vortices \cite{LagoudakisK2008}. Integer and fractional quantum Hall states are examples of quantum Hall fluids (QHFs). The distinction arises from an integer or fractional factor connecting the number of formed quantised vortices to a magnetic flux number associated with the applied field. The fractional factors present richer physics content than its integer cousin. These include the braiding statistics \cite{Laughlin1983} and the recently conjectured double boundary layer \cite{WiegmannP2013}. In the same vein, the edges of fractional QHFs, under small Zeeman splitting, supports spin textures \cite{HZi-Xiang2011}. The case of a QHF with the accompanying properties in graphene will be particularly interesting. Chiefly due to graphene unique magnetotransport properties \cite{MRabiu2012,LKayoung2011,OostingaJ2010}, unusual THz generation and amplification \cite{MRabiu2012b} and observation of anomalous quantum Hall states at room temperature \cite{DXu2009,BolotinI2009,DeanC2011}. Another important feature missed by conventional two dimensional semiconductor heterostructures, caused by edge reconstruction, is the unusual edge charge and spin propagation \cite{HZi-Xiang2011}. In graphene fluid, there is a parameter window in which the edge reconstruction can be avoided \cite{HZi-Xiang2011}. In practice, this seem to have paved the way for the experimental observations of charge and spin excitations at the edges of fractional quantum Hall liquids \cite{HZi-Xiang2011,ZhangY2013,STakei2014}. 

On the other hand, quantum interpretations of experimental measurements based on quasiparticle theory revealed that edge spin (spin-flip) excitations are more energetically favourable compare to edge charge (conserved-spin) excitations and even spin waves in the bulk \cite{RRoldan2010}. Ability to tune such spin excitations is very crucial in spintronics, which is witnessing an increasing interests through coherent spin dynamical properties. For instance, dissipationless spin superfluid transport realised on the platform of spintronics device could be very attractive, in terms of large data storage and information processing. Bulk spin superfluid transport has been reported \cite{STakei2014}. The case of spin-polarised fractional quantum Hall states have also been studied for spin reversed excitations \cite{MajunderD2014,WurstbauerU2011}. However, the main hindering factor towards realising a practical system will be the energy cost. Edge spin transport should be able to overcome the obstacles.

In this work, we study collective excitations concerning spin-vortex dynamics localised at Hall fluid edges. The fluid boundaries are modelled as smooth edges to capture the true double boundary layer property of fractional quantum Hall liquids. This is done in view of quantised Euler hydrodynamics theory for vorticity. The hydrodynamic approach is motivated by the fact that QHF is a manifestation of microscopic properties on the macroscopic scale. Hydrodynamic modes at finite frequencies emerge at long wavelength limit due to slow motion of quasiparticles. It is also based on its simplicity and ability to properly capture quasiparticles interactions. The new term arising from such interactions, called anomalous term \cite{WiegmannP2013b,AbanovA2013} has non-trivial consequence on transport properties, which is missed by microscopic theories \citep{HZi-Xiang2011,ZhangY2013}. The hydrodynamic theory have been used to correctly yield exact Laughlin states \cite{WiegmannP2013b} and bulk charge excitations \cite{AbanovA2013}. 

Microscopically, graphene has fourfold spin ($\uparrow, \downarrow$) and pseudo-spin ($K,K'$) degeneracy. Within the macroscopic picture, the system possesses four fluid components though only one momentum flux is measured. Each component belongs to the space $\{K,K'\}\otimes\{\uparrow, \downarrow\}$.

The remaining of the paper is organized as follows. In Section \ref{Sec:Theory}, a solution for vortex flow will be derived from quantized Euler hydrodynamic equation. We will calculate the edge density and edge spin excitations for filling factors within $0 \leq \nu \leq 1$ together. The universal similarities and differences in behaviour will be discussed in Section \ref{Sec:Section3}. We will conclude in Section \ref{Sec:Section4} highlighting possible applications of our results.

\section{Theoretical model}\label{Sec:Theory}

\subsection{Euler hydrodynamics of charge vortices}
The dynamics of incompressible vortex fluid is governed by Euler and continuity equations, 
\begin{equation}
 D_t\rho^{\alpha} = 0\quad \mbox{ and }\quad D_t{\bf u}^{\alpha} + \nabla {\rm p}^{\alpha}= 0.\label{Eq:EFluid}
\end{equation}
Where the material derivative $D_t \equiv \partial_t + {\bf u}\cdot {\bf \nabla}$ and ${\bf u}$ is macroscopic fluid velocity connected with the microscopic electron velocity, ${\bf v} = v_F{\bf k}/k$. ${\rm p}$ is the partial pressure per density and $\alpha$ is the fluid component index (K$\uparrow$, K$\downarrow$, K'$\uparrow$, K'$\downarrow$). Taking the curl of Eq.~(\ref{Eq:EFluid}), we get
\begin{equation}
  D_t\omega^{\alpha} = 0.\label{Eq:VFluid}
\end{equation}
Where the vorticity $\omega = {\bf \nabla}\times {\bf u}$. The continuity equation for the vortex, $ D_t\rho^{\alpha}_{\rm v} = 0$ also holds. The solution of Eq.~\ref{Eq:VFluid} have been obtained by Helmholtz and reported in \cite{Flutcher1999}. In a disk geometry having image vortices, the solution consists of point-like vortices of the form, 
\begin{equation}
  {\rm u}^{\alpha} =  {\rm i}\sum\frac{\Gamma^{\alpha}}{z^{\alpha} - z_i^{\alpha}(t)} + c.c+b.t,\label{Eq:Vortex0}
\end{equation}
where $c.c$ is the complex conjugate, $z = x + {\rm i} y$ and $u = u_x - {\rm i} u_y$. $b.t$ is the boundary term which includes drift velocity arising from a confining potential at fluid boundary. We have assumed a flow in which the strength of circulation, $\Gamma^{\alpha}_i$ ($=\Gamma^{\alpha}$) is both minimal and chiral such that in the thermodynamic limit, rotation can be compensated by the large number of vortices. This means, one can have Bohr-Sommerfeld phase-space quantization  of the circulations, $m_{\text{v}}\Gamma^{\alpha} = 2\pi\beta^{\alpha}\hbar$. Where $m_{\text{v}}$ is the vortex inertia and $\beta^{\alpha}$ is an integer or fraction. The equation of motion for vortices is such that 
\begin{equation}
  \text{ v}_i^{\alpha}(t) = {\rm i}\sum_{i\neq j}\frac{\Gamma^{\alpha}}{z_i^{\alpha}(t) - z_j^{\alpha}(t)} + c.c+b.t. \label{Eq:Vortex}
\end{equation}

Writing {\rm v} interms of {\rm u} using the identities $\pi\delta(r) \equiv \bar{\partial }(1/z)$ and   $\sum_{i\neq j}[2/(z-z_i)(z_i-z_j)] \equiv [\sum(1/(z-z_i)]^2 - \sum[1/(z-z_i)]^2$ , it is straight forward to show that~\cite{WiegmannP2013b} $  {\rm v} = {\rm u}+({\rm i}\Gamma/2)\partial_z{\rm log}\rho_{\rm v}$. We can recast the expression as
\begin{equation}
  \rho_{\rm v}{\rm v} = \rho_{\rm v}{\rm u}+\frac{\Lambda_{\nu}}{4}\Delta{\rm u} + \rho_{\rm v} {\rm v}_D. \label{Eq:EFlux}
\end{equation}
using the commutation relation $[{\rm u},\rho]={\rm i}(\hbar/m_{\rm v})\nabla\rho$. Where ${\rm v}_D$ ($=e\ell_B^2\nabla V/\hbar$) is the drift velocity of vortices due to the boundary terms and $\Lambda_{\nu} = 1 - 1/\beta + 2m_{max}$. With $m_{max}$ the angular momentum of vortices localised near the boundary.

\subsection{Gilbert-Landau-Lifshitz equation of spin vortices}
The equation of motion for spin-vortices having a unit magnitisation, {\bf m} is best decribed by the Gilbert-Landau-Lifshitz equation (GLLE) \cite{ThieleAA1973,HuberDL1982}
\begin{equation}
	\bar{D}_t{\bf m}_{i}+(\gamma/m_0)\varepsilon_{ijk}{\bf m}_j\partial_{{\bf m}_k}\mathcal{H}_s+\varepsilon_{ijk}{\bf m}_j\tilde{D}_t{\bf m}_k=0.\label{Eq:GLLE1}
\end{equation}
Where the material derivatives are $\bar{D}_t = \partial_t+{\bf j}\cdot\nabla$ and $\tilde{D}_t= \alpha\partial_t + \beta{\bf j}\cdot\nabla$. {\bf j} is the spin-vortex current and $\mathcal{H}_s$ is the spin Hamiltonian \cite{ThieleAA1973}. $m_0$ is the magnitude of the magnetic moment per area.  $\alpha$ is the Gilbert damping constant, which quantifies the macroscopic response of spin angular momentum polarised along z-axis. It takes small values for temperatures far below room temperature \citep{HuberDL1982}. $\gamma$ is the gyromagnetic ratio and $\beta$ is the non-adiabatic effect. The magnetisation is projected out-of-plane, for this reason one can map the in-plane coordinate ($(r,\phi)$) onto a 3D unit sphere ($\Theta(r),\Phi(\phi)$). That is, ${\bf m} = (sin\Theta(r)sin\Phi(\phi), sin\Theta(r)cos\Phi(\phi), cos\Theta(r))$. We are concerned with the dynamics in the vicinity of of an equilibrium fluid that may result in the deformation of the excitation spectrum. Equation (\ref{Eq:GLLE1}) can be transformed into non-Newtonian force of interactions between vortices \citep{HuberDL1982}
\begin{equation}
F_{\mu} = \varepsilon_{\mu\nu}N_0(U_{\nu}-J_{\nu}) + \delta_{\mu\nu}D_{0}(\alpha U_{\nu}-J_{\nu}).\label{Eq:GLLE2}
\end{equation}
governing the drift of the spin-vortices. $U$ and $J$ are the spin velocity and current, respectively. Assuming a vanishing dissipation dyadic, $D_0=0$, the number $N_0$ is expressed as
\begin{equation}
	N_0 = m_0\int{\rm d}^2r\left[\nabla\Theta\times\nabla\Phi\right]sin\Theta.\label{Eq:TopNum}
\end{equation}
Equation (\ref{Eq:TopNum}) is solved using the constraint $\Phi = \Gamma\phi+\pi/2$ together with the quantisation rule $[\Phi,\nabla\Theta]=\pm i\beta\hbar$. These yield, $N_0=\pm2\hbar\beta(m_0+\nabla\rho_{\rm v})$. Assuming for the moment that $J<<U$, one obtains the the spin velocity in a plane $U = F/N_0$ or in terms of energy of vortex-vortex interactions \citep{HuberDL1982}, $W=-2\pi(m_0\beta^2\hbar^2J_s/\gamma)\sum_i\log|z-z_i|$.
\begin{equation}
	U = \pm\frac{m_0\beta J_s}{\gamma(m_0\pm\nabla\rho_{\rm v})}{\rm u}.\label{Eq:SpinVel0}
\end{equation}

For the intrinsic spin degree of freedom, one can redefine the charge and spin densities as $2\rho_c = \rho_{\uparrow}+\rho_{\downarrow}$, $2\rho_s = \rho_{\uparrow}-\rho_{\downarrow}$ and velocities as ${\rm u} = {\rm u}_{\uparrow}+{\rm u}_{\downarrow}$, ${\rm U} = {\rm u}_{\uparrow}-{\rm u}_{\downarrow}$. Equation (\ref{Eq:EFlux}) is transformed into the velocity field
\begin{equation}
  {\rm v} = {\rm u} + \frac{\Lambda_{\nu}}{4(\rho_c^2-\rho_s^2)}\left(\rho_c\Delta{\rm u}-\rho_s\Delta{\rm U}\right) + {\rm v}_{D,c}, \label{Eq:SpinVel1}
\end{equation}
which can be simplified further in terms of {\rm u} using Eq.~(\ref{Eq:SpinVel0}).

\subsection{Edge velocity and model of boundary term} 
In real space and at the boundary, Eq.~(\ref{Eq:Vortex0}) assumes the form
\[
	{\rm u}(r)\Big|_{\partial\Omega}=-\sum_i\frac{g\Gamma(r-r_i)}{|r-r_i|^2}+\sum_i\frac{g\Gamma(r-r'_i)}{|r-r'_i|^2}
\]
or in momentum space
\[
	{\rm u}(q)\Big|_{\partial\Omega}=-\sum_ig\Gamma\left(\frac{r^2-r_i^2}{2\pi r}\right)\int_{r=R}{\rm d}^2r\frac{e^{{\rm i}qr cos\theta}}{|r-r_i|^2}.
\]
Following standard elementary calculations, this simplifies to give
\begin{equation}
	{\rm u}(q)\Big|_{\partial\Omega}=-\sum_ig\Gamma\frac{\varepsilon_{\ell}\Gamma}{R}\left(\frac{r_i}{R}\right)^{|\ell|}{\rm i}^{\ell}J_{\ell}(qR)e^{i\ell\theta_q}.\label{Eq:BoundVel}
\end{equation}
$\theta_q$ is the angle between {\bf q} and {\bf R} introduced after performing the $\theta$ integration.

The spin edge dispersion is calculated using the continuity equations defined in the bulk and on the edge of the fluid, as
\begin{equation}
	{\rm i}\omega \rho_{c,B} = \nabla{\rm P}\Theta(R-r),\,
	{\rm i}\omega \rho_{c,E} = {\rm P}\delta(R-r).\label{Eq:BulkEdge}
\end{equation}
Where the momentum flux is ${\rm P}=\rho_c{\rm v}+\rho_s{\rm V}$. ${\rm V}$ is the spin-vortex velocity which differs from the charge velocity-vortex ${\rm v}$ by a proportionality constant. $\delta(x)$ and $\Theta(x)$ are delta and step functions of $x$. Combining Eqs.~(\ref{Eq:BulkEdge}) and using $\rho_c=\rho_{c,B}+\rho_{c,E}$, we get
\begin{equation}
	\omega = [{\rm v}+\mathcal{S}_z{\rm V}](\delta(r-R)-{\rm i}\Theta(r-R)\nabla[\log({\rm v}+\mathcal{S}_z{\rm V}) + \log\rho_c]
\end{equation}
integrating over ${\rm d}^2r$ on both sides utilising $\int\,\Theta(r-R)\nabla_r^{(n)}f(r)dr = -\nabla_r^{(n-1)}f(R)$ yields
\begin{equation}
	\omega R = \left[(1+\mu\mathcal{S}_z)(2+\log\rho_c)\right]{\rm v}\Big|_{\partial\Omega}.\label{Eq:BulkEdge2}
\end{equation}
Where the spin-charge competition effect $\mu=U/u=V/v=\xi/\bar{\xi}$ with $\bar{\xi} = (\hbar/2\pi m_{\rm v})R_0^2$ and $\mathcal{S}_z=\rho_s/\rho_c$. The velocity for vortices localised at the boundary $\partial\Omega$ is computed from Eq.~(\ref{Eq:SpinVel1}). In momentum space, this can be express as 
\begin{eqnarray}
  {\rm v}(q)\Big|_{\partial\Omega} &=&\sum_{i,\ell} \bar{\xi}\beta R\Big[\left(\frac{r_i}{R}\right)^{|\ell|} - \frac{q\Lambda_{\nu}}{2\rho_cR}\frac{(1-\mu\mathcal{S}_z)}{(1-\mathcal{S}_z^2)}\times\nonumber\\
		&&\left(\frac{r_i}{R}\right)^{|\ell-1|}\Big]K_{\ell}(qR) + {\rm v}_D\delta_{\ell,1}J_{\ell}(qR). \label{Eq:EFluxQuan4}
\end{eqnarray}

We now introduce the smooth edge model of FQHE fluid, which gives us qualitative understanding of spin-vortex dynamics at the edges of the system. To model the edges, we consider a sufficiently large but finite droplet consisting of $N_{\rm v}$ vortices occupying a disk with edges at equilibrium radius $R_0 = \ell_B\sqrt{N/\pi\rho_0}$.  The smoothness is caused by clustering of vortices at the edges and subsequently performing small cyclotron oscillations. It is possible to have dipole moment within the boundary layer. Modulations of charges in the layer can be related to the change in local radius, $\delta R$. For sufficiently smooth edges, this can be expanded in Fourier series $\delta R(\varphi) = R_0\sum_{m}b_{m}Exp({\rm i}m\varphi) + c.c$ for all possible angular momenta excitations. $m$ is an integer. Where the dimensionless coefficient $b_{\ell}$ is a geometric (deformations) parameter. It measures the smoothness (strength) of the edge and varied within $0 < b_{m} < 1$. After some algebra and simplifications, the velocity Eq.~(\ref{Eq:EFluxQuan4}) together with Eq.~(\ref{Eq:BulkEdge2}) now takes the form
\begin{eqnarray}
  \omega &=&\sum_{i,\ell} \bar{\xi}\beta \Big[\left(\frac{r_i}{R_0}\right)^{|\ell|}F_{1,n}-\frac{q\Lambda_{\nu}}{2\rho_cR}\frac{(1-\mu\mathcal{S}_z)}{(1-\mathcal{S}_z^2)}\nonumber\\
		&&\times F_{2,n}\left(\frac{r_i}{R_0}\right)^{|\ell-1|}\Big]K_{\ell}(qR_0)\nonumber\\
		&& + {\rm v}_D(-1)^nF_{1,n}\delta_{\ell,1}J_{\ell}(qR_0). \label{Eq:EFluxQuan5}
\end{eqnarray}
Where we have used the modified Bessel function relation $K_{\ell}(\Lambda x) = \Lambda^{-\ell}\sum_n[(\Lambda^2-1)x/2]^nK_{\ell+n}(x)$ and  the Bessel function $J_{\ell}(\Lambda x) = \Lambda^{-\ell}\sum_n[(1-\Lambda^2)x/2]^nJ_{\ell+n}(x)$ with $\Lambda = \sum_{}b_mExp({\rm i}m\varphi) + c.c$. In Eq.~(\ref{Eq:EFluxQuan5}), we have integrated over $\varphi$ to get $F_{m,n}(b_m) = \int {\rm d}\varphi \,\Lambda^{m-1}(\Lambda^2+1)^n$. 

So far, we have not provided physical interpretations to $\beta$. In the context of fractional quantum Hall effect fluids, it was realised as the Laughlin filling fractions for a single component fluid \cite{WiegmannP2013b}. For the case of graphene, which is a multicomponent fluid, we interpret $\beta(=1/\nu_G)$ as the Laughlin-Halperin filling factors \cite{GoerbigMO2012}. Where $\nu_G$ is the graphene filling fraction calculated from $\nu_G=\sum_{i=1}^4\nu_i$. $i$ is the component index. The component fractions are computed from \cite{GoerbigMO2012} $N_{\phi}=m_{\alpha}\rho_{\rm v,\alpha}+\sum_{\beta\neq \alpha}n_{\alpha\beta}\rho_{\rm v,\beta}$. The intra and inter component factors are well chosen to yield a particular $\nu_G$ value. By defining the intrinsic spin polarisation as $\mathcal{S}_z = \nu_{\uparrow K}+\nu_{\uparrow K'}-\nu_{\downarrow K}-\nu_{\downarrow K}$, each $\nu_G$ will yield a particular value for the $\mathcal{S}_z$. Previous equations equally hold for the pseudo-spin polarisation also defined as $\tilde{\mathcal{S}}_z =\rho_{ps}/\rho_c$ in terms of charge and $\tilde{\mathcal{S}}_z = \nu_{\uparrow K}+\nu_{\downarrow K}-\nu_{\uparrow K'}-\nu_{\downarrow K'}$ in terms of component index factors.

We are interested in simulating Eq.~(\ref{Eq:EFluxQuan5}) for various values of $\nu_G$, $\mathcal{S}_z$ ($\tilde{\mathcal{S}}_z$), $b_m$ and $\mu$.

\section{Results and Discussions}\label{Sec:Section3}
\begin{figure}[h!]
	\begin{center}
	\includegraphics[scale=0.24]{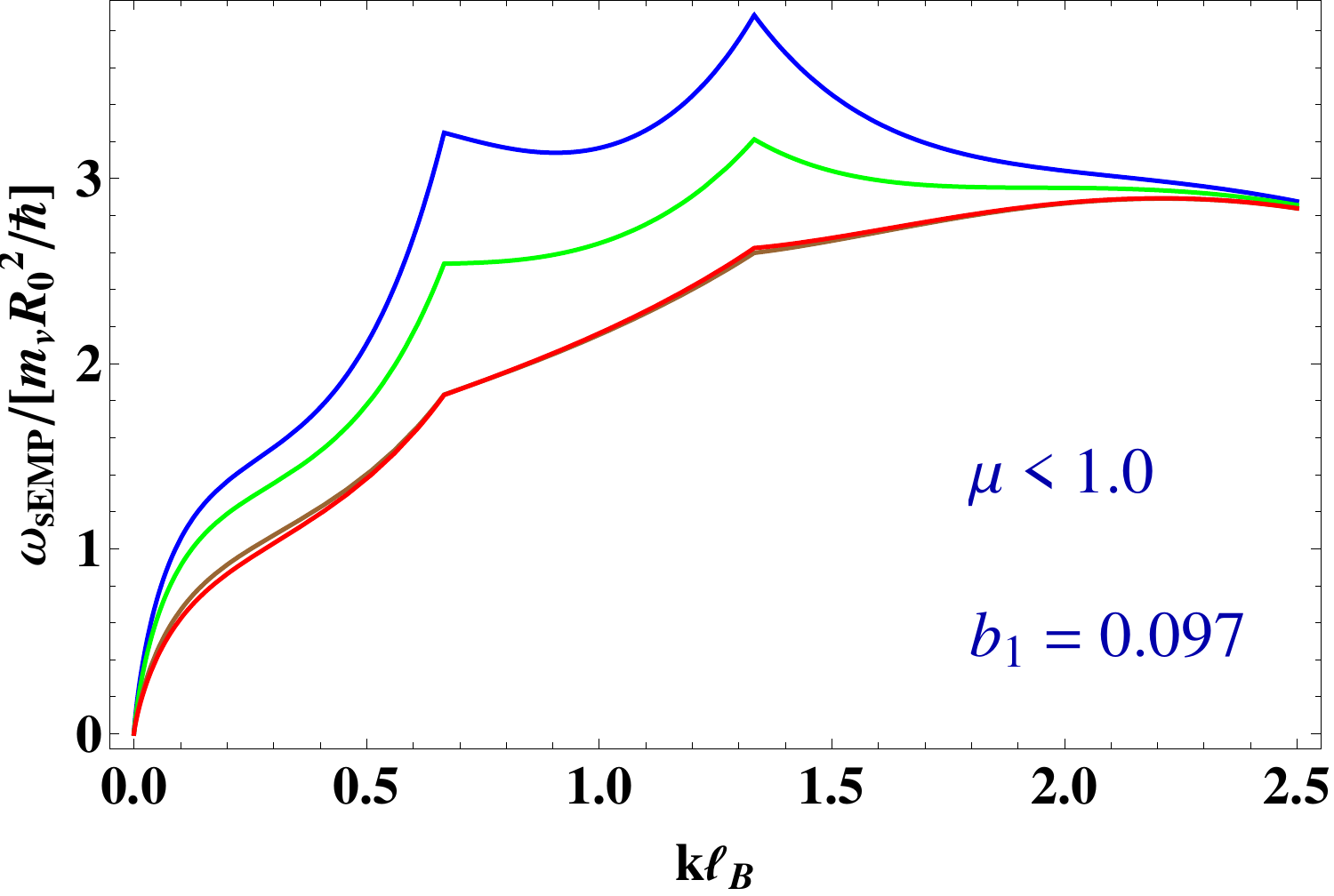}
	\includegraphics[scale=0.24]{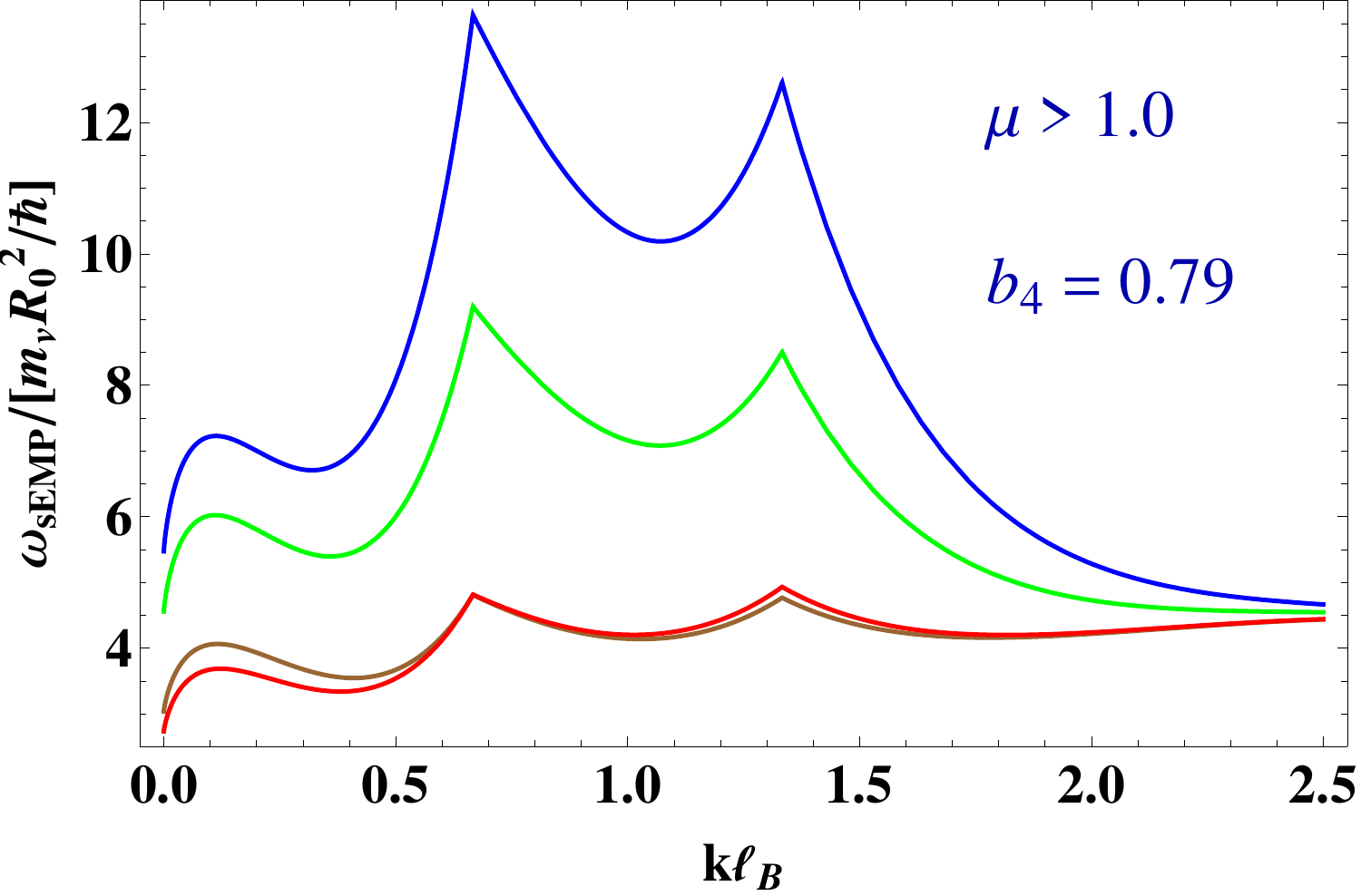}
	\includegraphics[scale=0.24]{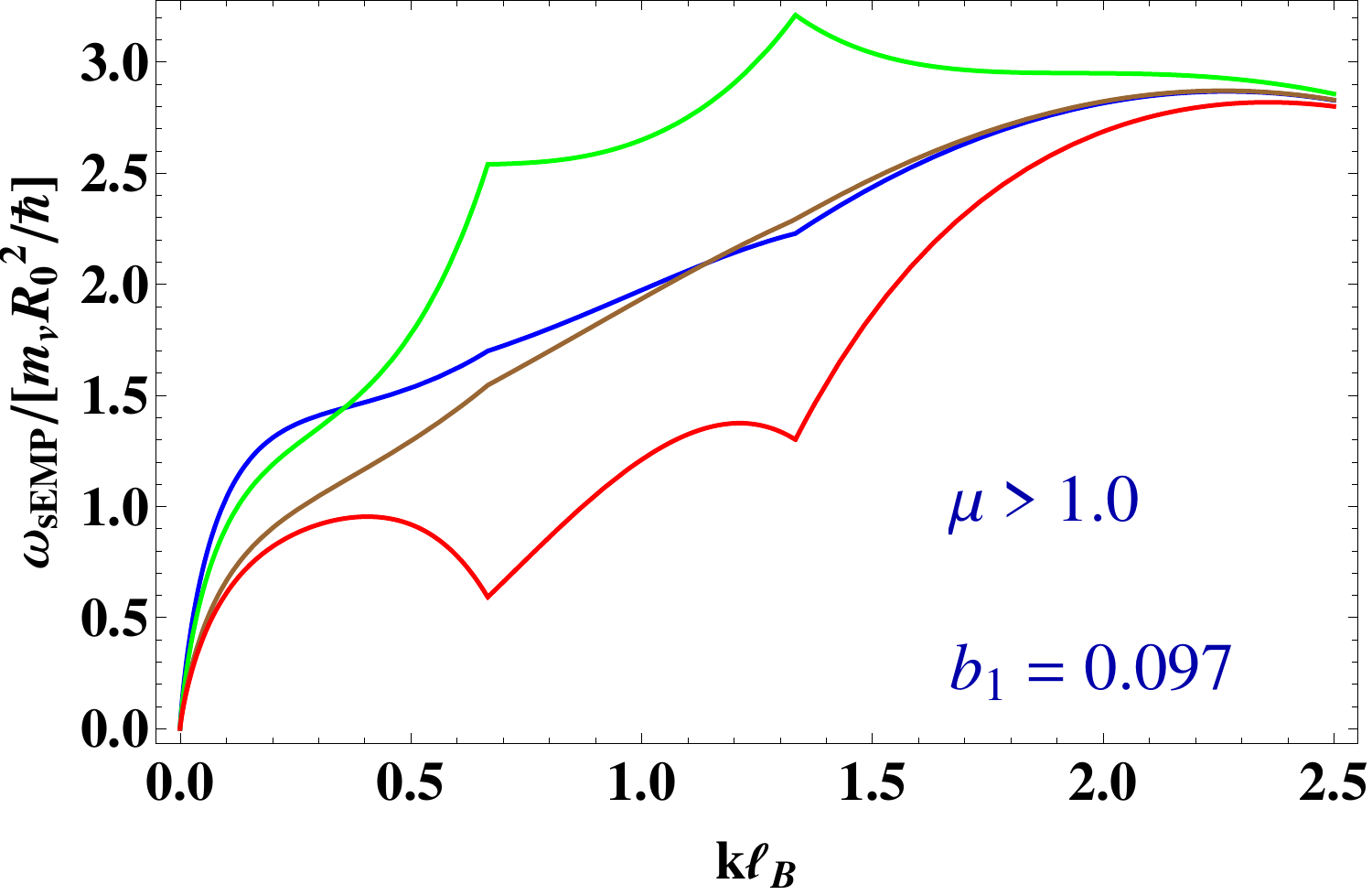}
	\includegraphics[scale=0.24]{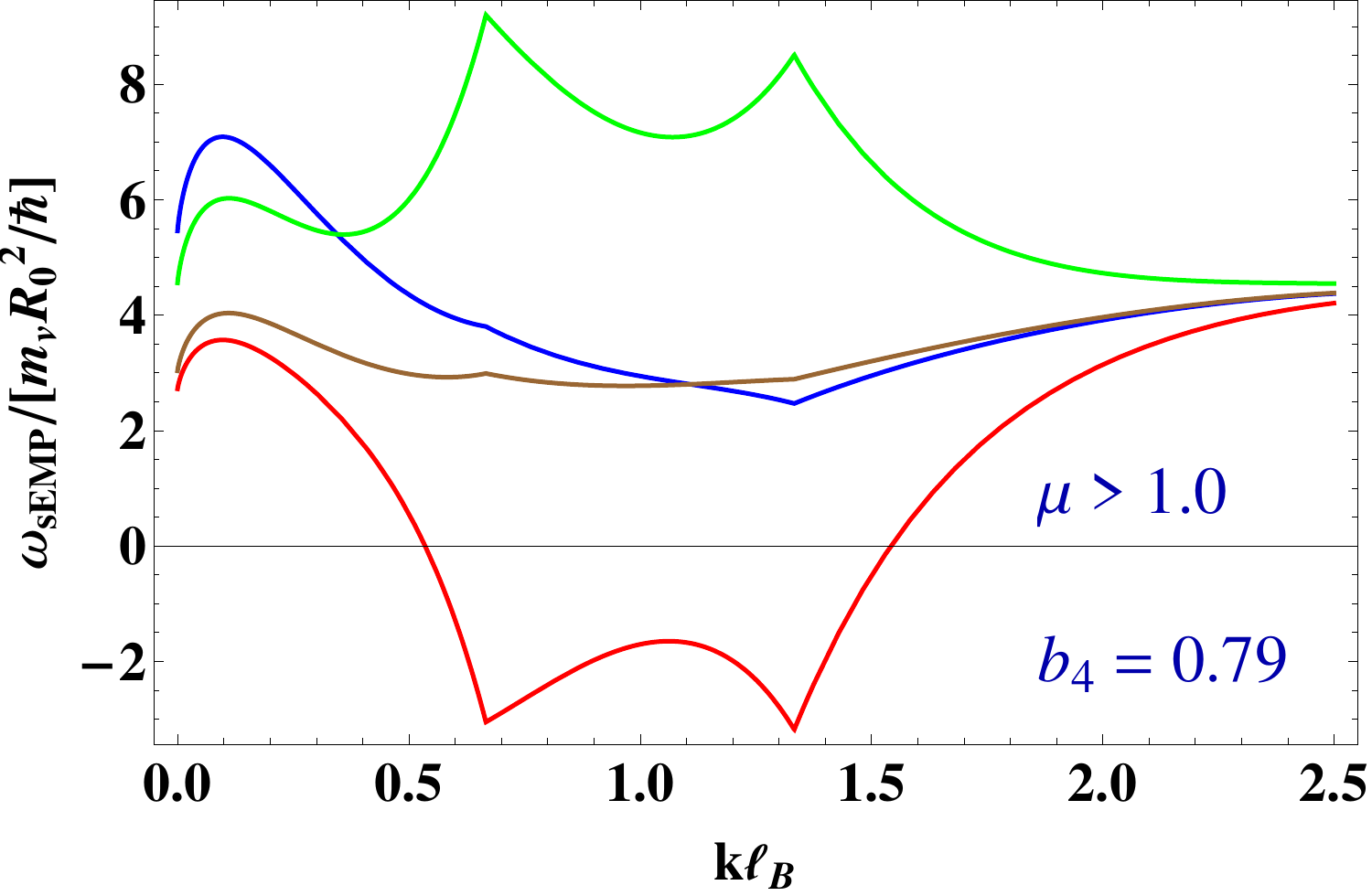}
	\caption{Spin excitation dispertions of (Top) $\mu>1$ and (Top) $\mu<1$ for spin, $\mathcal{S}_z$ polarisations. $\beta = 3$ (blue), $\beta = 5/3$ (green), $\beta = 5/2$ (brown), $\beta = 3/2$ (red).}\label{Fig:SEMP7}
	\end{center}
\end{figure}

\begin{figure}[h!]
	\begin{center}
	\includegraphics[scale=0.24]{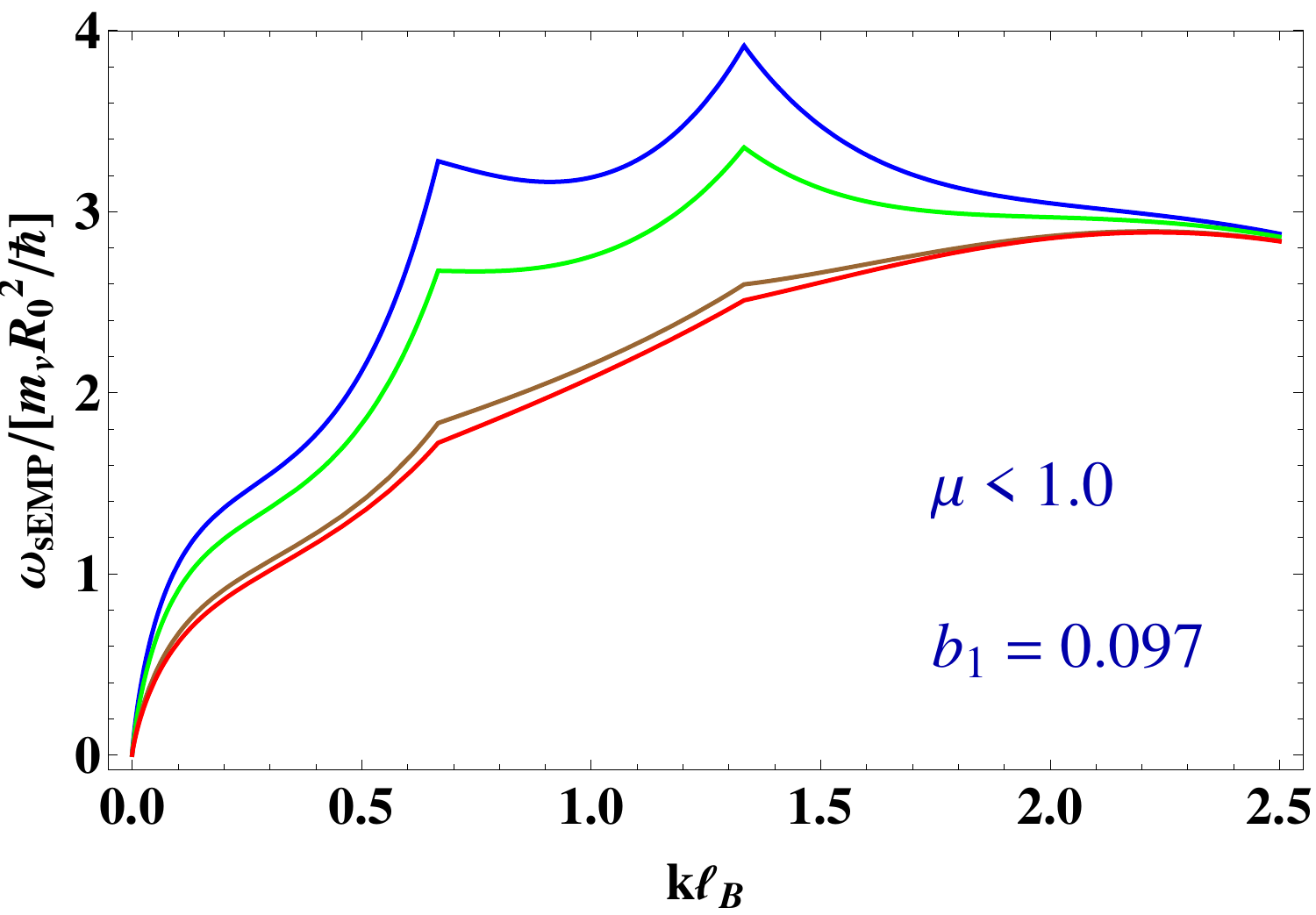}
	\includegraphics[scale=0.24]{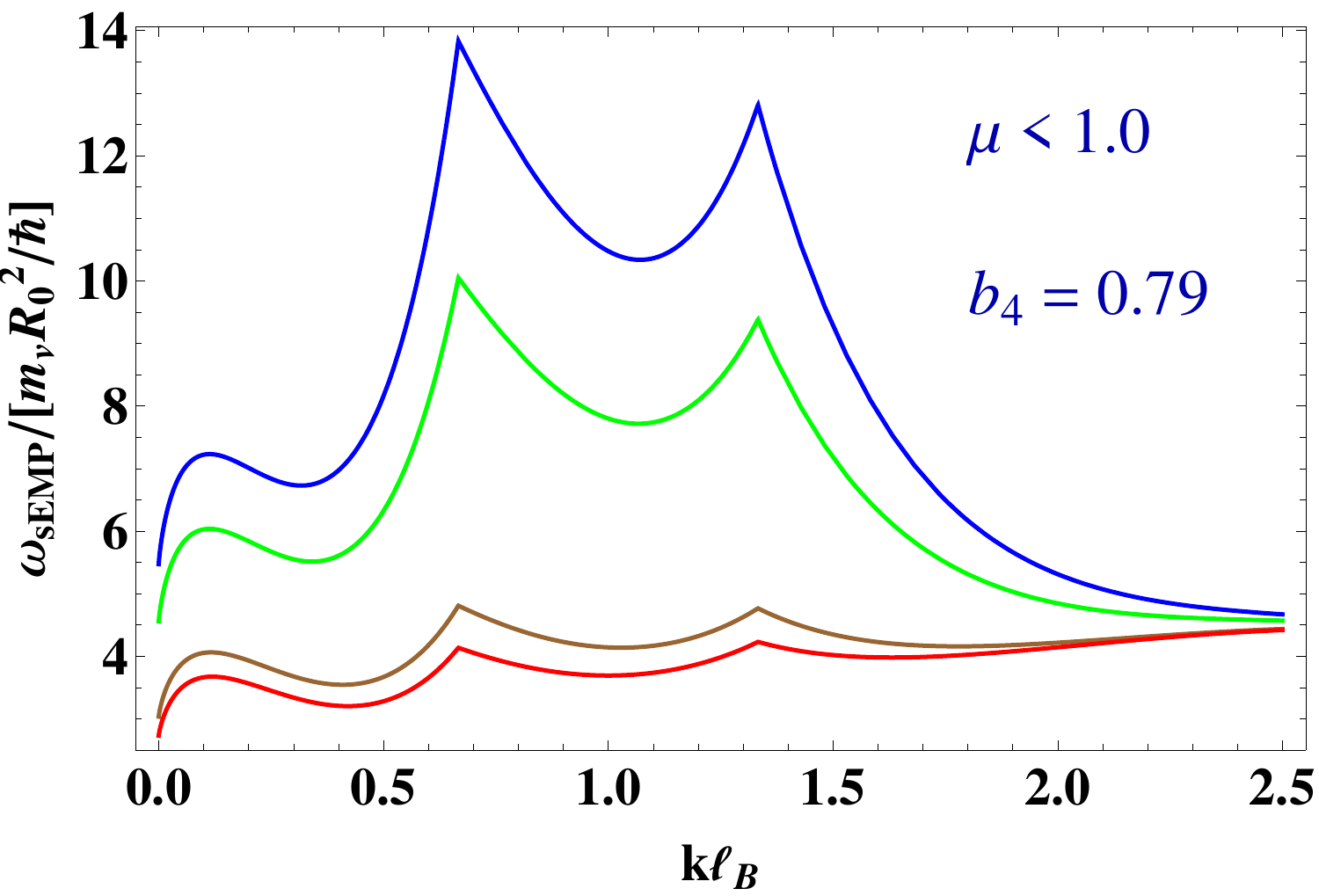}
	\includegraphics[scale=0.24]{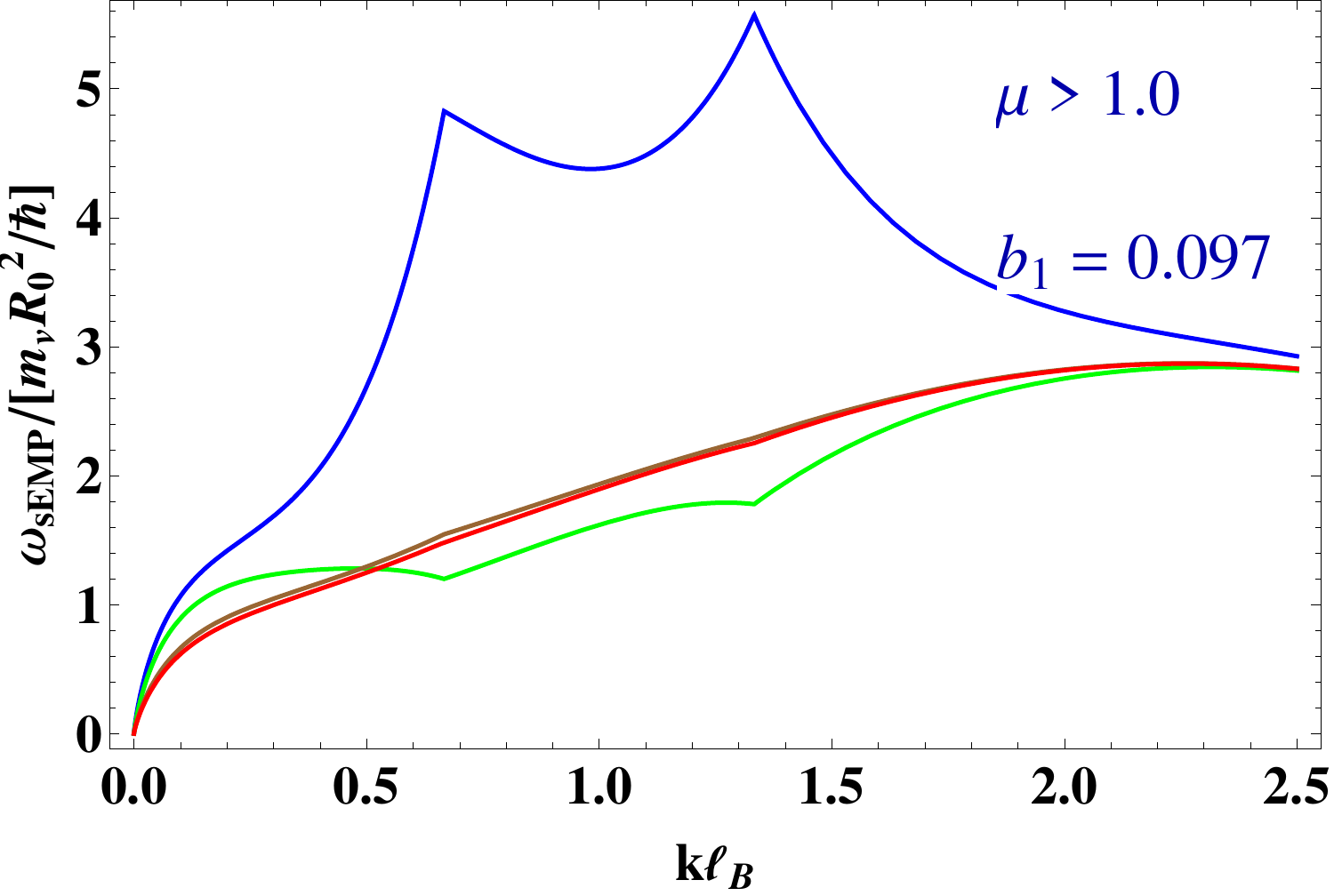}
	\includegraphics[scale=0.24]{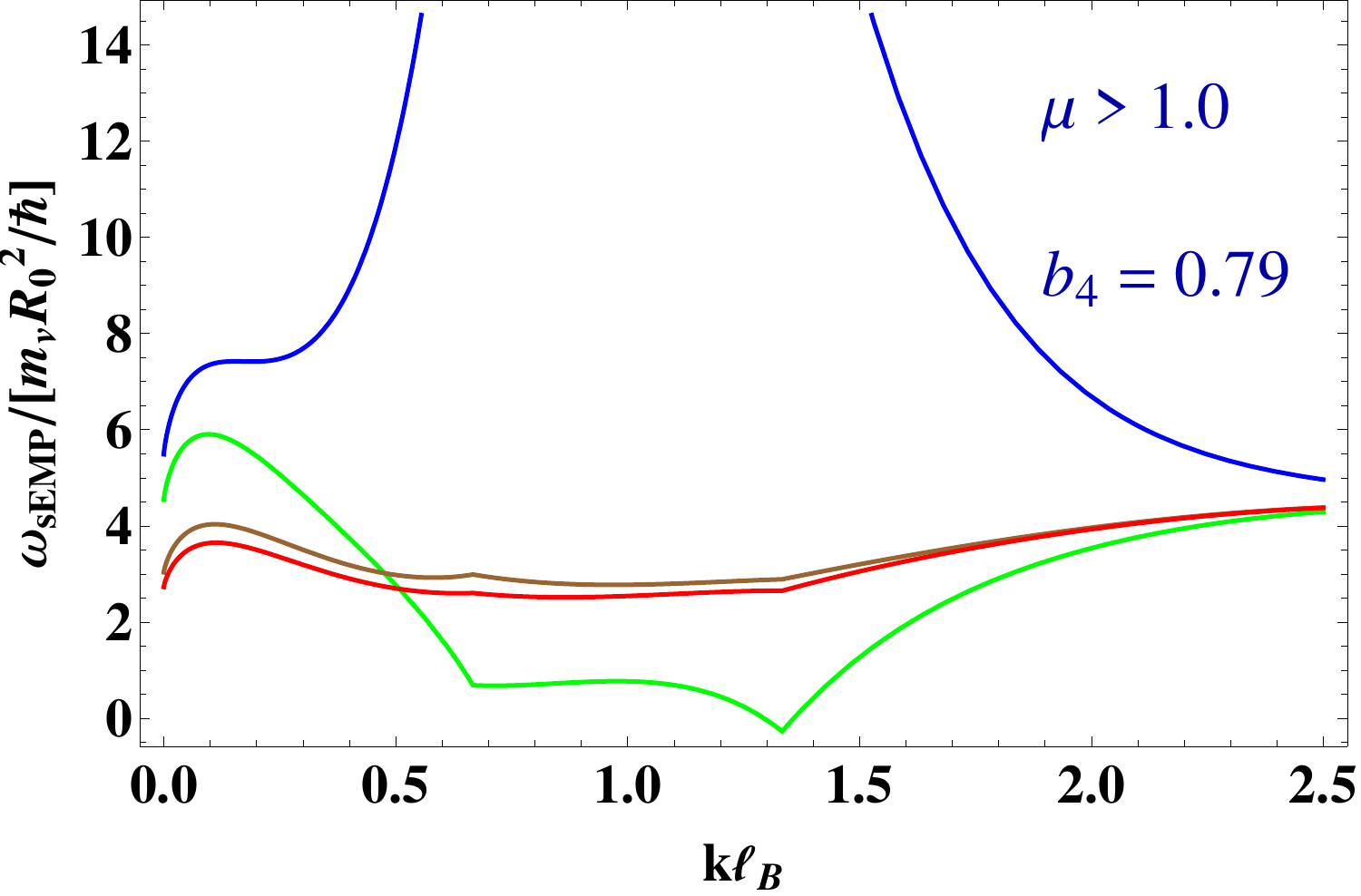}
	\caption{Spin excitation dispertions of (Top) $\mu>1$ and (Top) $\mu<1$ for spin, $\tilde{\mathcal{S}}_z$ polarisations. $\beta = 3$ (blue), $\beta = 5/3$ (green), $\beta = 5/2$ (brown), $\beta = 3/2$ (red).}\label{Fig:SEMP8}
	\end{center}
\end{figure}

In Fig. \ref{Fig:SEMP7} and Fig. \ref{Fig:SEMP8}, we observed the spin-vortex excitations for $\mu > 1$, $\mu < 1$ and deformation parameters; $b_1=0.097$, $b_4=0.79$. 

In Fig.~\ref{Fig:SEMP7} (Top) for $\mathcal{S}_z=(1/3,0,1/5,2/3)/2$ and $\tilde{\mathcal{S}}_z = 0$. When $\mu <1$, quasiparticle charge dynamics completely masked the spin dynamics. What is observed is the one resembling edge magnetoplasmons excitations. However, the excitation curves depicts resonance peaks originating from the boundary layer edges. Except the strong dependence on deformations, all filling factors have similar behavior. In Fig.~\ref{Fig:SEMP7} (Buttom) for the $\mu > 1$, the spin degree of freedom is very responsive making excitations appear unusual compared to reported observations \cite{ZhangY2013}. This can be seen in all the dispersions curves. Remarkably, an unusual behavior of coherent spin flipping are observed. On one hand, both $\beta=5/3$ and $\beta=3$ indicate partial spin polarisations (spin flip up and down), with respect to boundary edges. The $\beta = 5/2$ is unexpectedly, remain robust. However, the $\beta = 3/2$ have complete reorientation of spins. On the other hand, what it means is that, there is a fraction of the spins pointing one direction at the outer edge and the other fraction pointing the opposite direction. Also, all spins point up at one edge and down at the second edge. The behavior allows the system to have pure spin Hall liquid behavior which is very useful in spintronic devices for data storage and information processing. 

In the pseudo-spin polarisation sector; $\mathcal{S}_z=0$ and  $\tilde{\mathcal{S}}_z=(-1/3,2/5,1/5,2/9)/2$. When $\mu >1$, excitations similar to those of the Fig.~\ref{Fig:SEMP7} (Top) are obtained. This is shown in Fig.~\ref{Fig:SEMP8} (Top). However, an interesting and unique spin-vortex dynamics resembling the case of Fig.~\ref{Fig:SEMP7} (Buttom) occurs. In Fig.~\ref{Fig:SEMP8} (Buttom), the spin-vortex flipping mechanisms for the pseudo-spin polarisations are quite different from the case of intrinsic spin polarisations. The robust state turn out to be the usual Laughlin state, $\beta=3$ as oppose to $\beta=5/2$. In general, boundary deformation causes more vortices to have their spins polarised. Less (more) energy is required to create $\beta=3$ ($\beta= 5/3$) excitations as indicated by the large difference in the observations for $b_1=0.097$ and $b_4=0.79$ graphs.

\begin{figure}[h!]
	\begin{center}
	\includegraphics[scale=0.2]{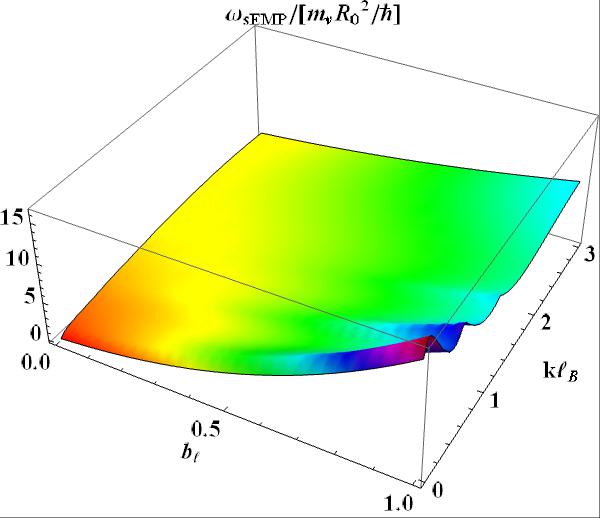}
	\includegraphics[scale=0.2]{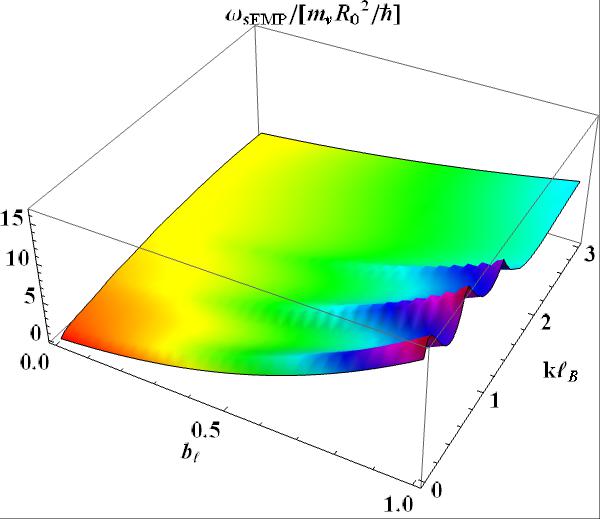}
	\end{center}
	\caption{Behaviour of sEMP excitations at varying edge deformation $b_{\ell}$, $\beta=3/2$ and $\mathcal{S}_z = 2/3$. (Left) $\mu>1$ and (Right) $\mu<1$.}\label{Fig:SEMP9}
\end{figure}

\begin{figure}[h!]
	\begin{center}
	\includegraphics[scale=0.2]{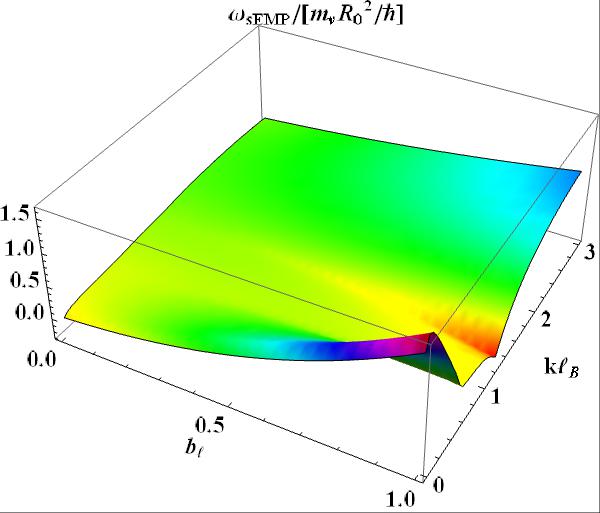}
	\includegraphics[scale=0.2]{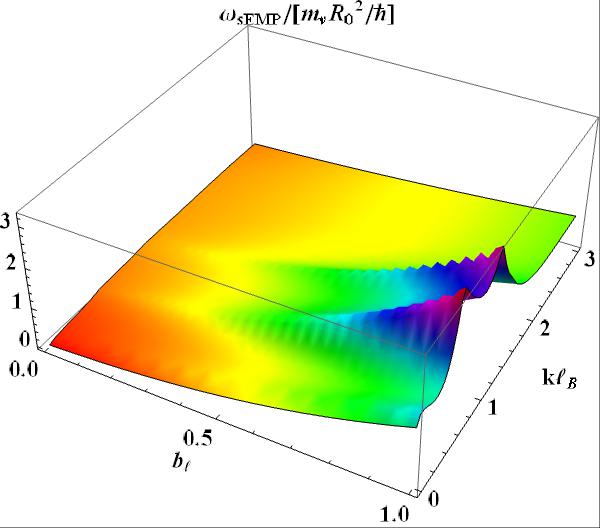}
	\caption{Behaviour of sEMP excitations at varying edge deformation $b_{\ell}$, $\beta=3/2$ and $\mu>1$. (Left) $\mathcal{S}_z = 0$, $\tilde{\mathcal{S}}_z = 2/3$ and (Right) $\mathcal{S}_z = 2/9$, $\tilde{\mathcal{S}}_z =0$.}\label{Fig:SEMP10}
	\end{center}
\end{figure}

A three dimensional plot showing the effects of varying deformations on the low-energy spin -vortex excitations is shown in Fig.~\ref{Fig:SEMP9}. The pseudo-spin polarisation corresponding to $\beta=3/2$ is used. It is immediately clear how the deformation parameter influences the vortex excitations when the spin-charge competition parameter is tune from right to left. Oscillations gradually appears at relatively high deformations. This means that plasmonic signals with constant amplitude are generated and the phenomenon is ideal for device applications. Similarly, for a constant $\mu$ but varying polarisation from pseudo-spin to spin polarisations, there is a total spin flip from up to down at both boundaries. This is depicted in Fig.~\ref{Fig:SEMP10}. Again, the controllability of the spin degree of freedom can be utilised in spintronic devices.
	
\section{Conclusions}\label{Sec:Section4}
In conclusion, we have demonstrated detailed theoretical studies of spin-vortex dynamics, using quantised Euler and Gilbert-Landau-Lifshitz equations. Two excitation peaks show up depicting the double boundary character of FQHE. In a physically relevant parameter space, some filling factors exhibit full and partial spin flipping at fluid boundaries. An inconsequential robustness is demonstrated by the $\beta=1/3$ and $\beta=5/2$ for pseudo-spin and spin polarisations, respectively. On the other hand, the other filling fractions show an unexpected evolution to partial or full polarisations. This is achieved by tuning both deformation and spin-charge parameters.

A natural direction to extend this studies is the edge excitations of spin-textured (Skyrmions) of the fractional quantum Hall fluid. A motivation was already presented for the integer case \cite{ZhangY2013}.

\begin{acknowledgments} 
Rabiu will like to thank International Center for Theoretical Physics (ICTP)-Italy for hospitality and provision of travel grants to conduct part of the work at the center in Trieste.
\end{acknowledgments}



\begin{thebibliography}{9}
\bibitem{Laughlin1983} Laughlin, R. B., "Anomalous quantum Hall effect: an incompressible quantum fluid with fractionally charged excitations", Physical Review Letters {\bf 50} (18) 1395, (1983).

\bibitem{PAlexandr2012} Pikalov, Alexandr A. and Dmitrii V. Fil, "Graphene bilayer structures with superfluid magnetoexcitons", Nanoscale Research Letters {\bf 7}(1), pp.1-9 (2012).

\bibitem{LagoudakisK2008} Lagoudakis, K. G., Wouters, M., Richard, M., Baas A., Carusotto, I., André, R., Dang, L.S. and Deveaud-Plédran, B., "Quantized vortices in an exciton–polariton condensate", Nature Physics {\bf 4}(9), pp.706-710 (2008).

\bibitem{WiegmannP2013} Wiegmann, P., "Nonlinear hydrodynamics and fractionally quantized solitons at the fractional quantum Hall edge", Physical Review Letters {\bf 108}(20), 206810 (2012).

\bibitem{HZi-Xiang2011} Zi-Xiang H., Bhatt,R. N.,  Xin W., and Yang, K., "Realizing universal edge properties in graphene fractional quantum Hall liquids", Physical review Letters {\bf 107}(23), 236806 (2011).

\bibitem{MRabiu2012} Rabiu, M., Mensah, S.Y. and Abukari, S.S, "General Scattering Mechanism and Transport in Graphene", Graphene {\bf 2}(01), 49 (2013).

\bibitem{LKayoung2011}  Kayoung L., Seyoung K., Points, M. S., Beechem, T. E., Taisuke O. and Tutuc, E., "Magnetotransport properties of quasi-free-standing epitaxial graphene bilayer on SiC: Evidence for Bernal stacking", Nano Letters {\bf 11}(9), 3624-3628 (2011).

\bibitem{OostingaJ2010} Oostinga, J.B., Benjamin S., Monica F. C. and Alberto F. M., "Magnetotransport through graphene nanoribbons." Physical Review B {\bf 81}(19) 193408, (2010).

\bibitem{MRabiu2012b} Rabiu, M., Mensah, S.Y., and Abukari, S.S, "Terahertz generation and amplification in graphene nanoribbons in multi-frequency electric fields." Physica E: Low-dimensional Systems and Nanostructures {\bf 61} pp.90-94 (2014).

\bibitem{DXu2009} Xu D., Ivan S., Fabian D., Adina L. and Andre, E.Y., "Fractional quantum Hall effect and insulating phase of Dirac electrons in graphene", Nature {\bf 462}(7270),  pp.192-195 (2009).

\bibitem{BolotinI2009}  Bolotin, K.I., Ghahari, F., Shulman, F.D., Stormer, H.L. and Kim, P., "Observation of the fractional quantum Hall effect in graphene", Nature {\bf 462}(7270), pp.196-199 (2009).

\bibitem{DeanC2011}  Dean, C. R., Young, A. F., Cadden-Zimansky, P., Wang	, L., Ren, H., Watanabe, K., Taniguchi, T., Kim, P., Hone, J. and Shepard, K.L., "Multicomponent fractional quantum Hall effect in graphene", Nature Physics {\bf 7}(9), pp.693-696 (2011).

\bibitem{ZhangY2013} Zhang, Y. and Kun Y., "Edge spin excitations and reconstructions of integer quantum Hall liquids", Physical Review B {\bf 87}(12), 125140 (2013).

\bibitem{STakei2014} Takei, S., Halperin, B.I., Yacoby, A. and Tserkovnyak, Y., "Superfluid spin transport through antiferromagnetic insulators."Physical Review B {\bf 90}(9), 094408 (2014).

\bibitem{RRoldan2010} Roldan, R., J-N. Fuchs, and M. O. Goerbig. "Spin-flip excitations, spin waves, and magnetoexcitons in graphene Landau levels at integer filling factors." Physical Review B {\bf 82}(20), 205418 (2010).

\bibitem{MajunderD2014} Majumder, D., and Sudhansu S. M., "Neutral collective modes in spin-polarized fractional quantum Hall states at filling factors 1/3, 2/5, 3/7, and 4/9". Physical Review B {\bf 90}(15), (2014) 155310.

\bibitem{WurstbauerU2011} Wurstbauer, U, Majumder, D., Mandal, S. S., Dujovne, I., Rhone, T. D., Dennis, B. S., Rigosi, A. F., Jain, J. K., Pinczuk, A., West, K. W., and Pfeiffer, L. N., "Observation of Nonconventional Spin Waves in Composite-Fermion Ferromagnets."Physical Review Letters {\bf 107}(6), 066804 (2011).

\bibitem{WiegmannP2013b} Wiegmann, P. B. "Hydrodynamics of Euler incompressible fluid and the fractional quantum Hall effect." Physical Review B {\bf 88}(24), 241305 (2013).

\bibitem{AbanovA2013} Abanov, Alexander G., "On the effective hydrodynamics of the fractional quantum Hall effect", Journal of Physics A: Mathematical and Theoretical {\bf 46}(29), 292001  (2013).

\bibitem{Flutcher1999} Flucher, Martin. Vortex motion in two dimensional hydrodynamics, Birkhäuser Basel (1999).

\bibitem{ThieleAA1973} Thiele, A. A., "Steady-state motion of magnetic domains", Physical Review Letters {\bf 30}(6), 230 (1973).

\bibitem{HuberDL1982} Huber, D. L., "Dynamics of spin vortices in two-dimensional planar magnets." Physical Review B {\bf 26}(7), 3758 (1982).

\bibitem{GoerbigMO2012} Goerbig, M. O. and Regnault, N., "Theoretical aspects of the fractional quantum Hall effect in graphene." Physica Scripta {\bf 146},  014017 (2012).
\end{thebibliography}
\end{document}